\documentclass{aa}
\usepackage{graphics}

\begin{document}
\title{Abundance and evolution of galaxy clusters in \\
cosmological models with massive neutrino
}
\author{ 
N.A. Arhipova
\inst{1}
\and
T. Kahniashvili
\inst{2,3}
\and
V.N. Lukash
\inst{1}
 }
\offprints{Arhipova N.A. \\
\email{arna@lukash.asc.rssi.ru}}

\institute{
Astro Space Center of P.N.Lebedev Physical Institute, 
Moscow, Russia
\and
Center for Plasma Astrophysics, Abastumani
Astrophysical Observatory, Tbilisi, Georgia
\and
Department of Physics and Astronomy,
Rutgers University,  Piscataway, NJ, USA
}
\date{Received  2 August 2001 / accepted  11 January 2002}
\abstract{
The time evolution of the number density of galaxy clusters 
and their mass and temperature functions are used to constrain 
cosmological parameters in the spatially flat dark matter models 
containing hot particles (massive neutrino) as well as cold and 
baryonic matter. 
We test the modified MDM ($\Lambda =0$) models with cosmic gravitational 
waves and show that they neither pass the cluster evolution 
test nor reproduce the observed height of the first acoustic 
peak in $\Delta T/T$ spectrum, and therefore should be ruled out. 
The models with a non-zero cosmological constant are in better agreement 
with observations. We estimate the free cosmological parameters in 
$\Lambda$MDM with a negligible abundance of gravitational waves, 
and find that within the parameter ranges $h\in (0.6, 0.7)$, 
$n\in (0.9, 1.1)$, $f_\nu\equiv\Omega_\nu /\Omega_m\in (0, 0.2)$, 
(i) the value of $\Omega_\Lambda$ is strongly affected by a small 
fraction of hot dark matter: $\;0.45 <\Omega_\Lambda <0.7\;$
($1\sigma$ CL), and 
(ii) the redshift evolution of galaxy clusters alone 
reveals the following explicit relation between $\Omega_\Lambda$ and 
$f_\nu$: $$\Omega_\Lambda +0.5f_\nu =0.65\pm 0.1.$$
This degeneracy is also expected in LSS tests (with a smaller error). 
The  present accuracy of observational data allows to bound 
the fraction of hot matter, $f_\nu\in (0, 0.2)$; the number of massive 
neutrino species remains undelimited, $N_\nu =1, 2, 3$. 

}
\titlerunning{Abundance and evolution of galaxy clusters}

\maketitle

\section{
Introduction
}
One of the most important problems of modern cosmology is
the formation of large-scale structure in the Universe (LSS).
The last decade progress in observations of LSS and cosmic microwave 
background anisotropy (CMBA) has allowed a productive comparison 
of theory with observations.
Any realistic cosmological model should be consistent with
LSS observational data in the range of scales from sub-Mpc 
(Gnedin 1998, Peackock 1997) up to the cosmological horizon 
(Smoot et al. 1992, Jaffe et al. 2001).

According to the standard theory of gravitational instability the 
observable LSS in 
the Universe has been formed by growth of small density inhomogeneities 
generated during the very early stages of cosmological expansion. Two main 
features of the primordial perturbation field -- 
the Gaussian character of linear density perturbations and the bottom-up 
LSS formation -- are confirmed by available astrophysical observations.  
Therefore, assuming that the initial conditions are given
one can derive the theoretical predictions for LSS formation in a dark 
matter (DM) model and then use statistical analysis and
observational data to test the viability of the considered models, 
i.e. to constrain the allowed range of principal cosmological parameters.

Nowadays, the most popular cosmological models are cold DM with 
non-zero cosmological constant ($\Lambda$CDM) (Peebles 1984, 
Kofman \& Starobinsky 1985) and mixed DM without (MDM) (Fang et al. 1984, 
Shafi \& Stecker 1984, Valdarnini \& Bonometto 1984, Lukash 1991, 
Lucchin et al. 1996) and with ($\Lambda$MDM) (Kahniashvili et al. 1996, 
Novosyadlyj et al. 1999b and refs. therein) cosmological constant, 
where MDM is in the form of non-baryonic cold (neutralino or 
hypothetical axions) and hot (massive neutrino) particles. 

The cosmological impact of non-baryonic collisionless matter depends on the free-streaming path of DM particles: at small scales 
($k>k_{FS}$) the perturbations smooth out, whereas at large scales ($k<k_{FS}$) they grow to form the gravitationally bounded objects of mass greater than
$M_{FS}$ (Novikov \& Zel'dovich 1975). 
Due to the various free-streaming pathes of cold and hot particles 
($k_{FS}^{(c)}\gg k_{FS}^{(\nu)}$) the growth rate of density perturbations is different in CDM and HDM components.
The linear perturbation power spectrum formed by redshift $z$, $P(k,z)$, depends strongly 
on the abundance of each component ($\Omega_c, \Omega_\nu$)
and is given by the production of the initial power spectrum, 
$P_0(k)$, the transfer function, $T(k,z)$, and the perturbation growth rate, 
$D(z)$ (e.g. Padmanabhan 1993). 
Both the transfer function and the growth factor are DM dependent and describe the evolution of initial density perturbations 
during expansion of the Universe (Zakharov 1979, Bardeen et al. 1986, Holtzman 1989, Eisenstein \& Hu 1998, Novosyadly et al. 1999a). 

 All cosmological models have been re-addressed after CMBA experimental 
detection (Bennett et al. 1996) to reveal their positive and negative 
features. 

Both $\Lambda$CDM and MDM models have met several difficulties.  

As far as $\Lambda$CDM models are concerned 
(Kofman et al. 1993, Eke et al. 1996, Liddle et al. 1996a,b, 
Primack \& Klypin 1996) they demand a high value 
of the cosmological constant ($\Omega_\Lambda \geq 0.7$). 
In this case $\Lambda$CDM is able to fit a set of LSS 
observational constraints whereas at small scales 
it overproduces the number of collapsed objects 
by a factor of 2 in comparison with the corresponding 
number of gravitationally bounded objects in galactic 
cataloges (Klypin et al. 1996, Gawiser 2000, Bond et al. 2000).

Regarding MDM models the difficulties are related to late 
galaxy (quasar) formation (Pogosyan \& Starobinsky 1995, Komberg et al. 1996) 
and too high numbers of obtained galaxy clusters (Ma 1996, 
Valdarnini et al. 1998, Gardini et al. 1999, Rahman \& Shandarin 2001). 
Standard MDM with one, two or three species of massive neutrino 
are ruled out at $2\sigma$ CL (Novosyadlyj et al. 1999b). 

A way to overcome the sMDM difficulties could be the 
consideration of decaying neutrinos (Bonometto \& Pierpaoli 1998),
or models with cosmic gravitational waves (CGW) (Arkhipova et al. 1999, 
Melchiorri et al. 1999), or a non-zero cosmological 
constant ($\Lambda$MDM) (Kahniashvili et al. 1996, Tegmark 1999). 
In this paper we consider the models with stable neutrinos.
 
An importance of fundamental CGW has been emphasized by 
Starobinsky (1979), Rubakov et al. (1982), Lukash \& Mikheeva (2000). 
Theoretical predictions for cluster abundance in MDM models with 
CGW has been presented by Ma (1996) for the red spectra of density 
perturbations ($n < 1$), and by Mikheeva et al.(2001)
for both blue ($n >1$) and red ($n <1$) scalar perturbation spectra. 
Although $\Lambda$CDM models with $\Omega_m \leq 0.3$ 
normalized by COBE 4-year data (Bunn \& White 1997) are consistent 
with the cluster number density test, in order to archive 
an agreement with CMBA and cluster abundance data in MDM models 
it is necessary to take into account the CGW contribution 
in the derived value of $\Delta T/T$ at $10^0$ angular scale. 
The latter is estimated by parameter T/S, 
the ratio of the tensor to the scalar mode contributions.

As an alternative to $\Lambda$CDM and MDM, the $\Lambda$MDM models 
have been considered by Valdarnini et al.(1998), 
Novosyadlyj et al.(1999b), Primack \& Gross (2000), Andres et al.(2000). 
The advantage of these models is in retaining the inflationary 
paradigm and the flat Harrison-Zel'dovich spectrum ($n=1$) with a smaller 
value (comparing to $\Lambda$CDM) of cosmological $\Lambda$-term. 
Other notable features of $\Lambda$MDM are the possibility of 
negligible CGW contribution (T/S =0) and a small fraction of 
hot particles: even $10\%$  of massive neutrino in matter content 
($\Omega_\nu/\Omega_m\sim 0.1$) could change the value of the 
cosmological constant, being however in good agreement with other
independent tests: CMB data (Hu et al. 2000), distant SNIa (Perlmutter et al.
1999), Hipparcos data (Feast \& Catchpole 1997), QSO lensing (Kochanek 1996).

In this paper we consider MDM (with CGW) and $\Lambda$MDM (with T/S=0) 
applying to these models the cluster abundance and evolution tests.
The evolution test constrains most efficiently the parameter $\Omega_m$, 
and hence $\Omega_\Lambda$ in spatially flat cosmological models 
(e.g. Bahcall et al. 1997, Donahue \&Voit 1999, Henry 2000). 
Our aim is to demonstrate how the presence of hot matter influences 
the estimation of $\Omega_\Lambda$, $h$, and other cosmological parameters.
Here, instead of doing an exact $\chi^2$ analysis, we rather look for 
the tendencies and correlations between cosmological parameters when 
introducing a fraction of massive neutrino in the Universe.
 
We describe our models in Section 2, and the galaxy cluster tests 
in Section 3. The results are discussed in Section 4, with the 
conclusions in Section 5. 

\section{Cosmological models with massive neutrino}

We assume that DM is given by mixture of CDM and HDM components 
in the flat background space. The free model parameters are: 
\begin{itemize}
\item
$\Omega_m$, the total matter density in the
Universe ($\Omega_m = 1-\Omega_\Lambda =\Omega_b+
\Omega_c+\Omega_\nu$,
the latter being density parameters of baryons, cold, and hot particles, 
respectively);\\ 
\item
$f_\nu\equiv\Omega_\nu /\Omega_m$, the fraction of hot DM;\\
\item
$N_\nu$, the number of massive neutrino species;\\
\item
$h\;$, the Hubble constant in units $100$ km s$^{-1}$Mpc$^{-1}$;\\
\item
$n\;$, the slope-index of post-inflationary density perturbation power 
spectrum.
\end{itemize}

The massive and massless neutrinos{\footnote{All neutrinos are active, the 
total number of neutrino species is three. Accordingly, $N_\nu = 1, 2, 3$.}} 
are described by the corresponding distribution functions which are 
evaluated from the Boltzmann-Vlasov collisionless equations. 
Cold particles are treated as a pressureless fluid ($p_C=0$). 
Baryons and photons are described as an ideal hydrodynamic fluid 
satisfying the Euler equations of motion.  Here we choose the fixed value 
for baryon density parameter, $\Omega_b=0.015/h^2$. 
All DM components interact with each other only gravitationally.

The calculation of the perturbation dynamics (the transfer functions) 
demands a joint solution of the described self-consistent set of 
equations. It can be done numerically (Seljak \& Zaldarriaga 1996) 
or analytically in long and short wave regions (Zakharov 1979, 
Bardeen et al. 1986, Holtzman 1989), the possibility of semi-analitical 
approximations of some transfer functions has been shown by 
Eisenstein \& Hu (1998), Novosyadlyj et al.(1999a).

Assuming the power low post-inflationary density perturbation
spectrum, $P_0(k) \infty k^n$, the final power spectrum of
total density perturbations
can be expressed as
\begin{equation}
P(k,z)=Ak^n T^2(k,z) 
\left [ {D(z) \over D(0)} \right]^2\;,
\end{equation}
where $A$ is the normalization constant, 
$T(k,z)$ is the total transfer function, $D(z)$ is the
growth factor{\footnote{The function $D(z)$ is given as the growth 
rate of linear density perturbations in the model without hot matter 
(with the same parameters $\Omega_m$, $h$, and $n$).}},
\begin{equation}
D(z)=\frac{g(\Omega_m(z))}{1+z}\;,
\end{equation}
and $g(\Omega_m(z))$ is the suppression coefficient 
(Kofman \& Starobinsky 1985). 

According to Carroll et al.(1992) the function $g(x)$ can be approximated as
\begin{equation}
g(x)=5x\left ( 2x^{\frac{4}{7}} + \frac{2+x(209-x)}{70}\right )^{-1}\;,
\label{supp}
\end{equation}
abd the current matter abundance $\Omega_m(z)$ is written as follows: 
\begin{equation}
\Omega_m(z)=\Omega_m \frac{(1+z)^3} {1- \Omega_m+(1+z)^3 \Omega_m}\;\;,\;\;\;\;
\Omega_m\equiv\Omega_m(0)\;.
\end{equation}
We use the transfer function approximations for sCDM (Bardeen et al. 1986) 
and $\Lambda$MDM (Eisenstein \& Hu 1998) models. 

\section{Cluster mass and temperature functions and 
the evolution test}

The mass function for the gravitationally bounded halos of mass greater 
than $M$ formed in the flat Universe by redshift $z$ is given by 
(Press \& Schechter 1974)
\begin{equation}
N(>M,z)=\int\limits_{M}^{\infty} n(M^{'},z) dM^{'}
\end{equation}
where $n(M,z)dM$ is the comoving number density of collapsed objects 
with masses lying in the interval ($M,M+dM$):
\begin{equation}
n(M,z) =
 \sqrt{\frac{2}{\pi}}\frac{\rho\delta_{c}}{M}
 \frac{1}{\sigma^{2}(R,z)} | \frac{d\sigma(R,z)}{dM} |
 e^{-\frac{\delta^{2}_{c}}{2\sigma^{2}(R,z)}}\;\;.
\label{N(>M)}
\end{equation}
$M=\frac 43\pi\rho R^{3}$,
$\rho$ is the mean matter density, and $\delta_c$ is the critical
density contrast for a linear overdensity able to collaps. 
The rms amplitude of density fluctuation in the spheres of 
radius $R$ at redshift $z$ is related to the power spectrum as
\begin{equation}
\sigma^{2}(R,z)= \frac{1}{2 \pi^{2}}\int\limits_{0}^{\infty} 
P(k,z)|W(kR)|^{2}k^{2}dk,
\end{equation}
where $W(kR)$ is a Fourier component of the top-hat window function:
$
W(x)=\frac{3}{x^3}(\sin{x}- x cos{x}).
$
We denote $\sigma_R\equiv\sigma(R,0)$ for $z=0$.

In the matter-dominated Universe $\delta_c=1.686$
(Eke et al. 1996, Viana \& Liddle 1996),
in the flat models with $\Lambda$-term $\delta_c$ depends weakly on 
the current matter abundance (Liddle et al. 1996a, Locas \& Hoffman 2000). 

The theoretical mass functions obtained with the help of the
Press-Schechter formalism (eqs.(5,6,7)) are in good agreement 
with other methods including numerical simulations 
(Efstathiou et al. 1988, Lacey \& Cole 1994, Gross et al. 1998,
Eke et al. 1998, Bode et al. 2000, Wu 2000).
Due to the exponential dependence of $n(M,0)$ on $\sigma_R$
(see eq.(6)) the  cluster number density is very sensitive to the value 
of $\sigma_8$~.

Rich galaxy clusters are strong X-rays sources characterised by 
the gas temperature. The physical mechanisms of clusters formation promp 
the relation between cluster mass and temperature, $T_g \infty M^{2/3}$, 
confirmed by numerical simulations (e.g. Navarro et al. 1995). 
The proportionality coefficient depends slightly on cosmological model 
(mainly, on the parameter $\Omega_m$). Here, we use the $T-M$ 
relation for isothermal gas given by Eke et al. (1996): 
\begin{eqnarray}
T_g= {7.75 \over \beta} \left ( {6.8 \over 5X+3} \right ) M_{15}^{{2 \over 3}} 
\left ( {\Omega_m \over \Omega_m(z)} \right )^{{1 \over 3}} 
\left ({\Delta_{cr} \over 178} \right )^{{1 \over 3}} (1+z)~\rm{Kev}
\label{Temmas}
\end{eqnarray}
where $\beta(\simeq 1)$ is the ratio of the galaxy kinetic energy
to the gas thermal energy, $X(\simeq 0.76)$ is the hydrogen mass fraction,
and the mass $M_{15}$ is given in units $10^{15} h^{-1} M_\odot$. 
The value $\Delta_{cr}$ is the ratio of a mean 
halo density (within the virial radius of collapsed object) to the 
critical density of the Universe at corresponding redshift. 
For $\Omega_m \leq 1$, $\Delta_{cr}$ can be derived analytically 
and approximated as $\Delta_{cr}=178 \Omega_m^{0.45}$ for $z=0$. 

As well as the cluster mass function $N(>M)\equiv N(>M,0)$, the 
X-cluster temperature function $N(>T)\equiv N(>T,0)$ is also
sensitive to $\sigma_R$.    
The comparison between the observed temperature (Henry \& Arnaud 1991) 
and mass (Bahcall \& Cen 1993) functions shows a good agreement with eq.(8) 
at $z=0$. Comparing with observational data both tests provide  
a powerful constraint on $\sigma$-value in different DM models 
(Valdarnini et al. 1998, Bahcall \& Fan 1998, Durrer \& Novosyadlyj 2001).  
On the other hand CMBA strongly depends on the density perturbation spectrum 
as well. The consistency with all these tests determines the model parameters 
(e.g. Bridle et al. 1999, Mikheeva et al. 2001).
 
 The cluster mass (temperature) function and its evolution in 
$\Lambda$CDM has been discussed in detail 
(e.g. Viana \& Liddle 1998, Bode et al. 2000, Henry 2000).
To achieve a better consistency with cluster evolution data
the value of $\Omega_m$ should be lower than that obtained from 
the cluster number density test at $z=0$.

The dependence of the cluster mass (temperature) function on $z$ 
originates due to different growth rate of density perturbations 
in DM models. It is reflected in the appearance of the suppression factor 
$g(\Omega_m(z))$ in models with $\Omega_\Lambda \neq 0$ (see eqs.(3,4)):
\begin{equation}
\sigma (R,z)=\frac{\sigma_R}{1+z}\frac{g(\Omega_m(z))}{g(\Omega_m)}\;. 
\end{equation}

Below, we consider the cosmological models containing massive 
neutrino. In this case $z$-dependence is also present in the 
transfer function due to the free-streaming of hot matter particles 
(the effective growth rate becomes scale dependent). For the mass 
estimation of rich galaxy clusters (see eq.(6)), the effect of
free-streaming is negligible.

\section{Results and Discussion}
\subsection{MDM models with non-zero tensor mode}

To test MDM models with a zero cosmological constant we follow 
the normalization procedure given by Arkhipova et al. (1999), 
Mikheeva et al. (2001). 
All models are normalized by $\sigma_8$ by the best fit present 
day cluster mass (Bahcall \& Cen 1993, Bahcall et al. 1997) and 
temperature (Henry \& Arnaud 1991, Donahue \& Voit 1999 ) functions.
To achieve an agreement with $\Delta T /T $ data (Bennet et al. 1996)
the contribution of cosmic gravitational waves is required. The value 
of the derived parameter T/S depends mainly on the spectral index $n$, the
abundance of hot matter $\Omega_\nu$, and the Hubble constant $h$. 

\begin{figure}
\resizebox{\hsize}{!}{\includegraphics{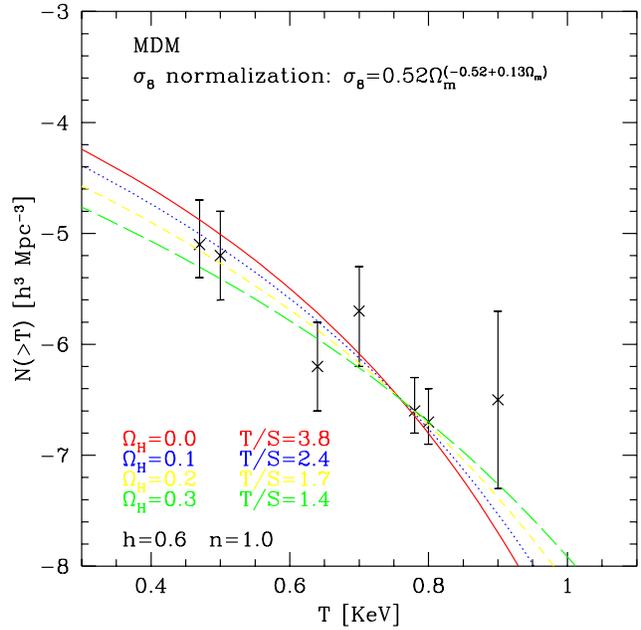}}
\caption{
The present day cluster abundance $N(>T)$  in MDM models normalized by $\sigma_8=0.52$ 
with $\Omega_\nu= 0, 0.1, 0.2, 0.3$ (solid, dot, dash, long-dash lines, resp.), $\Omega_b=0.015/h^2$, $h=0.6$, $n=1$. The needed T/S is shown. The data points by Henry \& Arnaud (1991).
}
\end{figure}

\begin{figure}
\resizebox{\hsize}{!}{\includegraphics{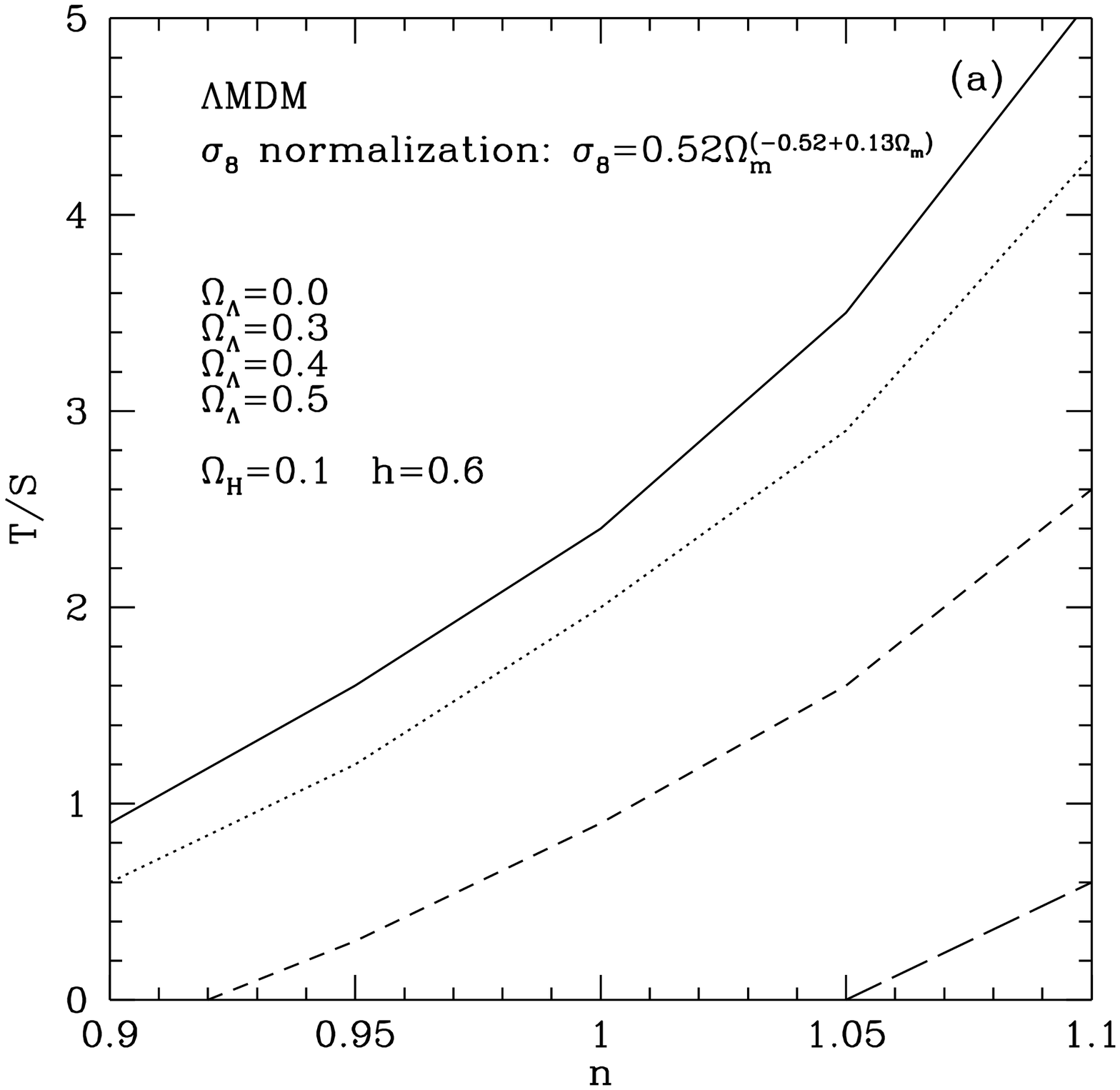}}
\resizebox{\hsize}{!}{\includegraphics{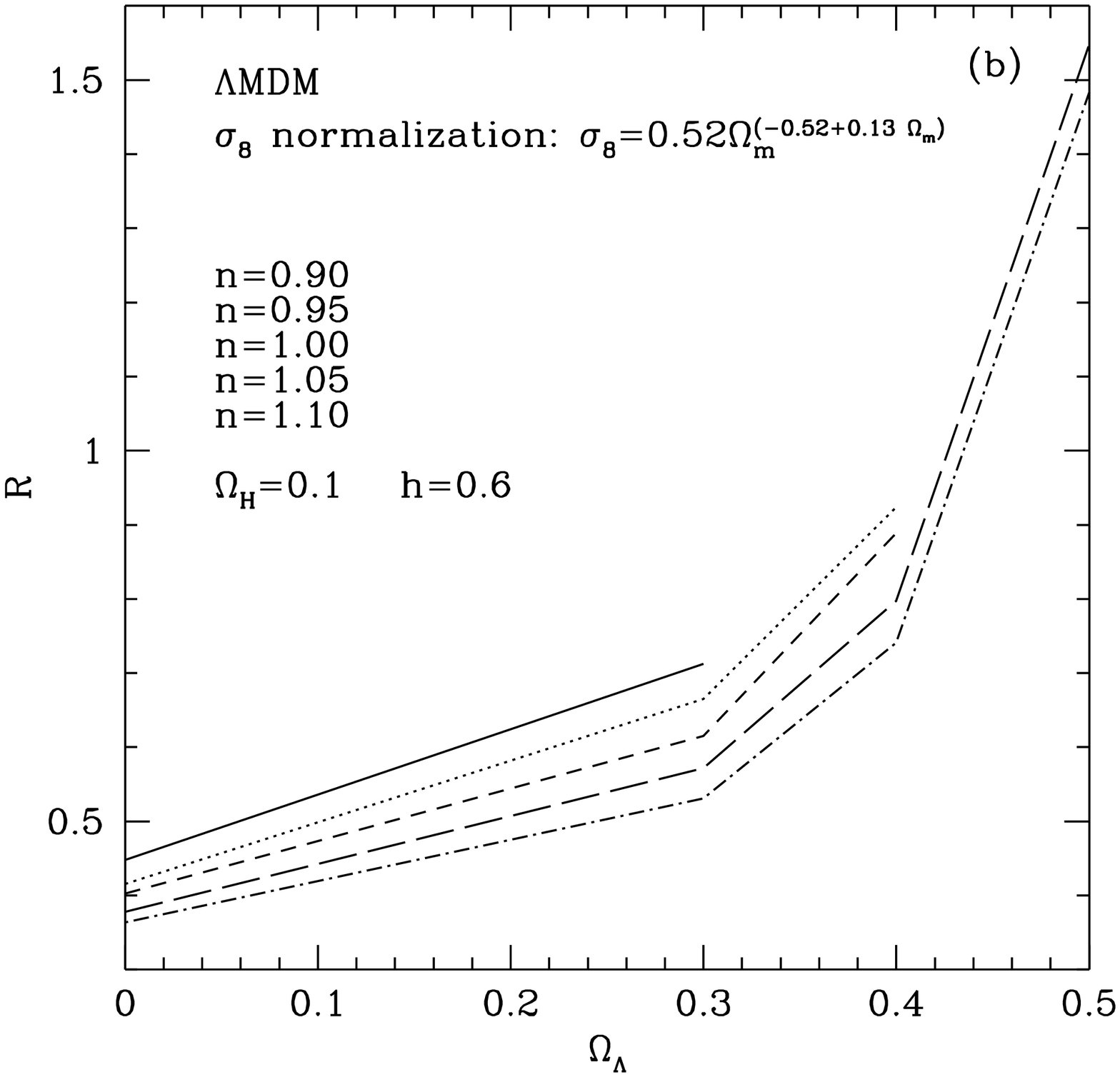}}
\caption{
The introduction of a cosmological constant in
MDM-CGW models normalized by 
$\sigma_8=0.52 \Omega_m^{(-0.52+0.13\Omega_m)}$ with 
$\Omega_\nu=0.1$, $\Omega_b= 0.015/h^2$, $h=0.6$:
(a) T/S as a function of $n$ for 
$\Omega_{\Lambda}=0, 0.3, 0.4, 0.5$ 
(solid, dot, dash, long-dash lines, resp.), and
(b) the height of the first acoustic peak, 
$R^{\frac 12}=\frac{\delta T_p}{70 \mu\rm{K}}$, 
as a function of $\Omega_{\Lambda}$ for 
$n=0.9, 0.95, 1, 1.05, 1.1$ 
(solid, dot, dashed, long-dashed, dashed-dot lines, resp.).} 

\end{figure}

\begin{figure}
\resizebox{\hsize}{!}{\includegraphics{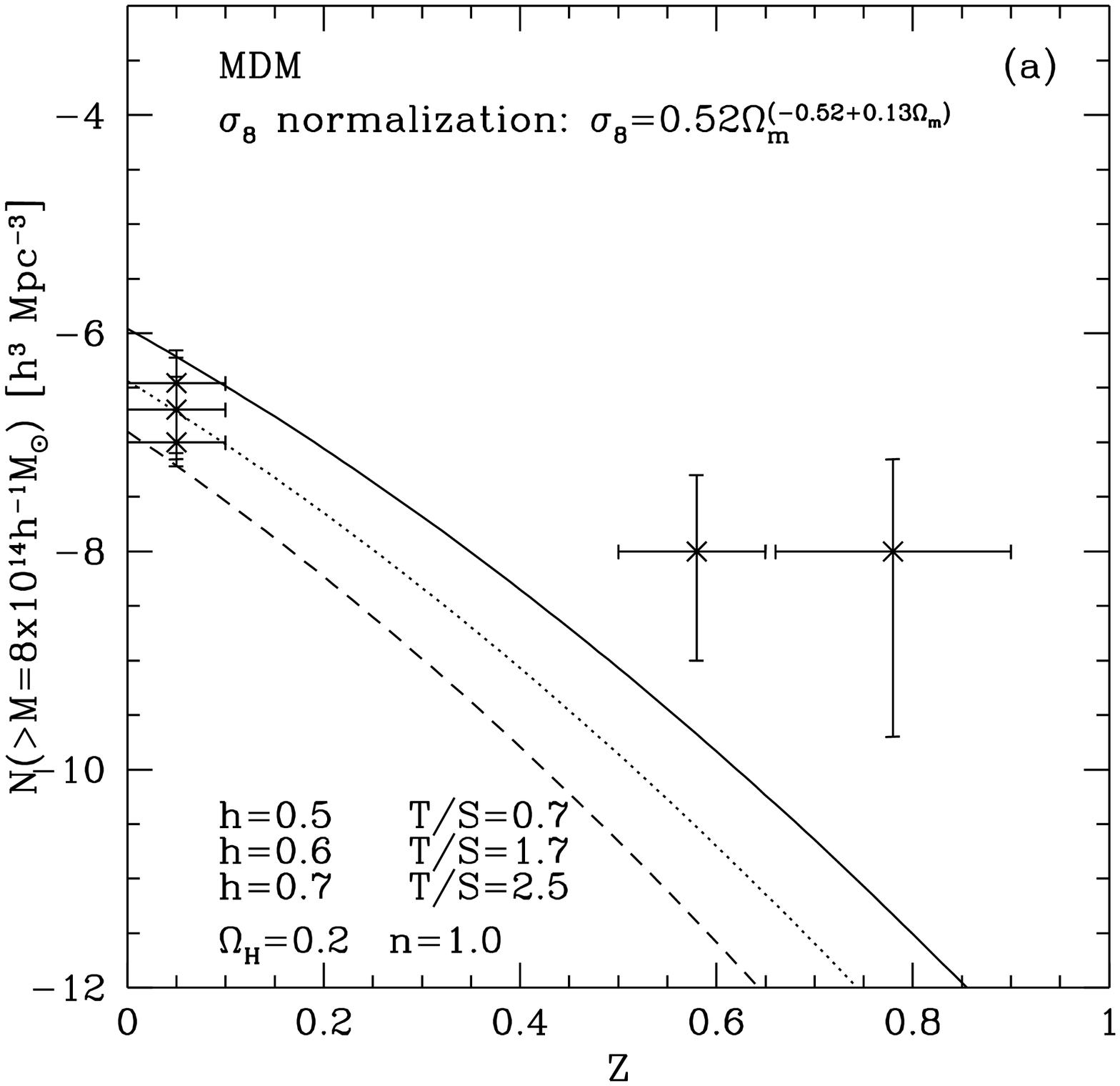}}
\resizebox{\hsize}{!}{\includegraphics{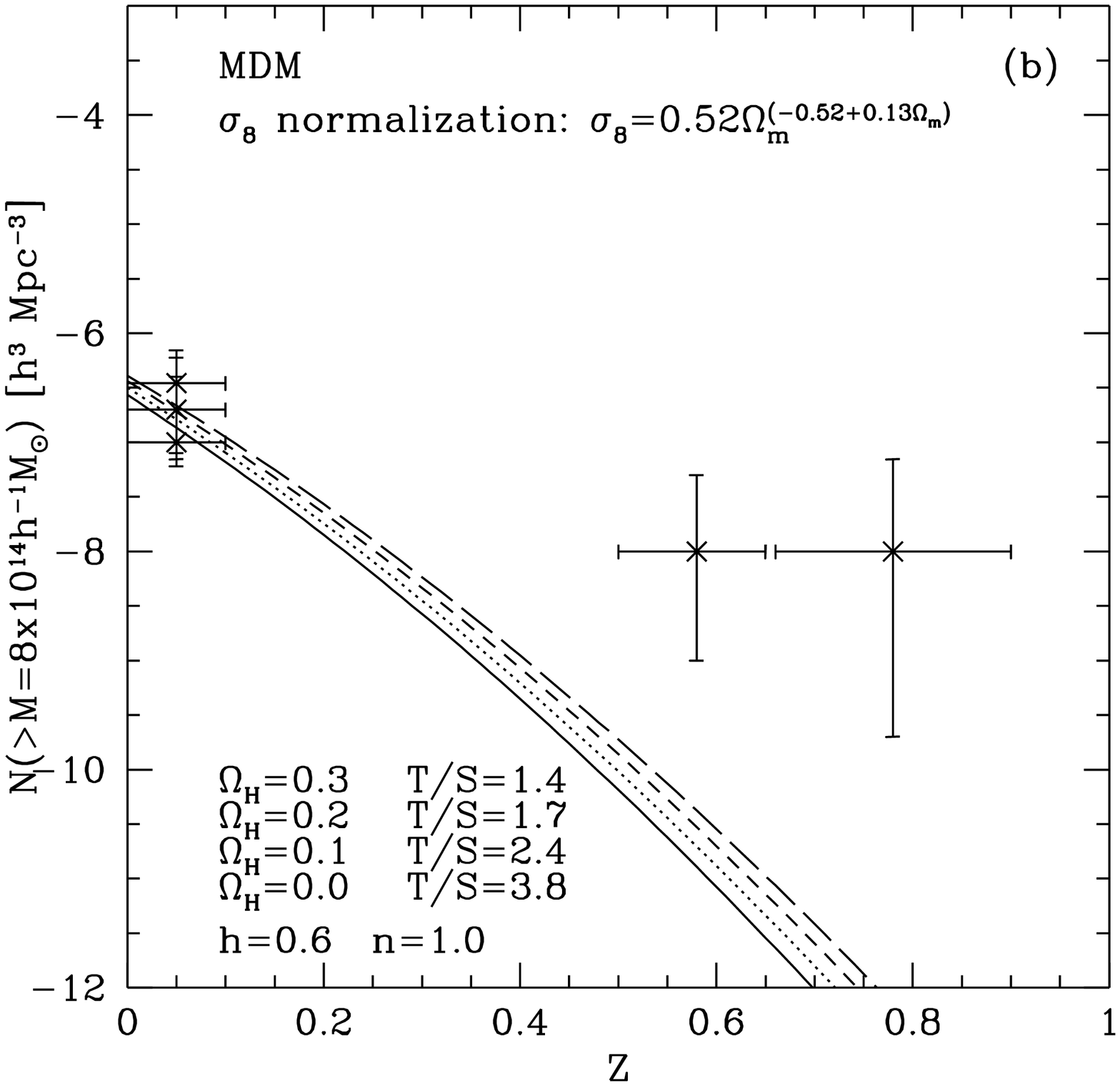}}      
\caption{
The cluster evolution $N(>M = 8 \cdot 10^{14} M_\odot, z)$ in MDM models 
normalized by $\sigma_8= 0.52$ with $n=1$, 
(a) $\Omega_\nu=0.2$,  
$h=0.5, 0.6, 0.7$ (solid, dot, dash lines, resp.), and
(b) $h=0.6$,  $\Omega_\nu=0, 0.1, 0.2, 0.3$  
(solid, dot, dash, long-dash lines, resp.).
The needed T/S is shown. The data points by Bahcall \& Fan (1998).
}
\end{figure}

Fig.1 shows the present-day cluster temperature functions $N(>T)$ 
for different values of $\Omega_\nu$ in MDM models with CGW normalized by 
$\sigma_8 \simeq 0.52$. The normalization does not depend on $\Omega_\nu$, 
therefore all curves cross each other at some fixed point corresponding 
to the mass $M$ in the sphere of radius $8h^{-1}$Mpc. 

A significant contribution of both CGW and massive neutrinos is 
needed to fit the large scale CMBA observational data. The consideration 
of the red spectra ($n <1$) cannot reduce T/S substantially.
E.g., in MDM models with fixed $\Omega_\nu=0.2$, $h=0.6$, and 
$n=0.9, 1, 1.1\;$: T/S=$0.6, 1.7, 3.6\;$ respectively, which strongly
suppresses the height of the first acoustic peak in $\Delta T/T$ spectrum 
(Mikheeva et al. 2001).

In this respect, higher values of $\Omega_\nu$ would be preferable 
for MDM models: an increase of $\Omega_\nu\geq 0.2$ reduces the 
parameter T/S, however the problem arises with small-scale clustering 
and the $Ly_\alpha$-cloud formation tests (Gnedin 1998). On the contrary, 
at low $\Omega_\nu < 0.2$ neither possible changes of $h$ nor models 
with three species of massive neutrino can decrease sufficiently a high 
contribution of CGW into CMBA at large angular scales. The latter 
obviously leads to a low height of the first acoustic peak inconsistent 
with the BOOMERANG observations (Netterfield et al. 2001). Fig. 2 
demonstrates that the introduction of even a small cosmological constant 
would drastically improve the situation with MDM-CGW model parameters.   

But let us return again to MDM models without $\Lambda$-term. 
Fig. 3 presents the evolution of galaxy clusters.  
None of the considered models can fit the data at high   
$z > 0.3$. We varied $h$ (Fig.~3a), $\Omega_\nu$ (Fig.~3b),
and $n$ (not shown). The evolution tracks are practically independent 
of the post-inflationary spectral index for $n\in (0.9, 1.1)$. 
Even neglecting the observational problems with $Ly_\alpha$-forest, 
the case with high $\Omega_\nu\geq 0.2$ cannot help to achieve an 
agreement with the evolution of galaxy clusters in MDM dominated 
models. All the models predict the number of galaxy clusters at high
$z\geq 0.3$ at least two orders of magnitude smaller than the observed one. 
This fact indicates that the evolution of density perturbations
should be slower than that found in MDM without a cosmological constant. 

\subsection{MDM models with non-zero cosmological constant}

Let us consider flat $\Lambda$MDM models without CGW normalized by 
the COBE 4-year data. This allows an explicit derivation of
$\sigma_8(\mbox{cmb})$ as a function of model parameters.
On the other hand, from observational abundance of galaxy clusters
it is possible to obtain $\sigma_8(\mbox{cl})$  as a function of $\Omega_m$, 
e.g. $\sigma_8(\mbox{cl})=0.52 \Omega_m^{-0.52 +0.13\Omega_m}$~ 
(Eke et al. 1996, Liddle et al. 1996b).
The coincidence between $\sigma_8(\mbox{cmb})$ and $\sigma_8(\mbox{cl})$ 
would limit the parameter $\Omega_\Lambda$ by $\sim 0.7$ in $\Lambda$CDM 
(Bahcall \& Fan 1998, Henry 2000) and $\sim 0.5-0.6$ in $\Lambda$MDM 
(Valdarnini et al 1998, Novosyadlyj et al. 1999b) models.

\begin{figure}
\resizebox{\hsize}{!}{\includegraphics{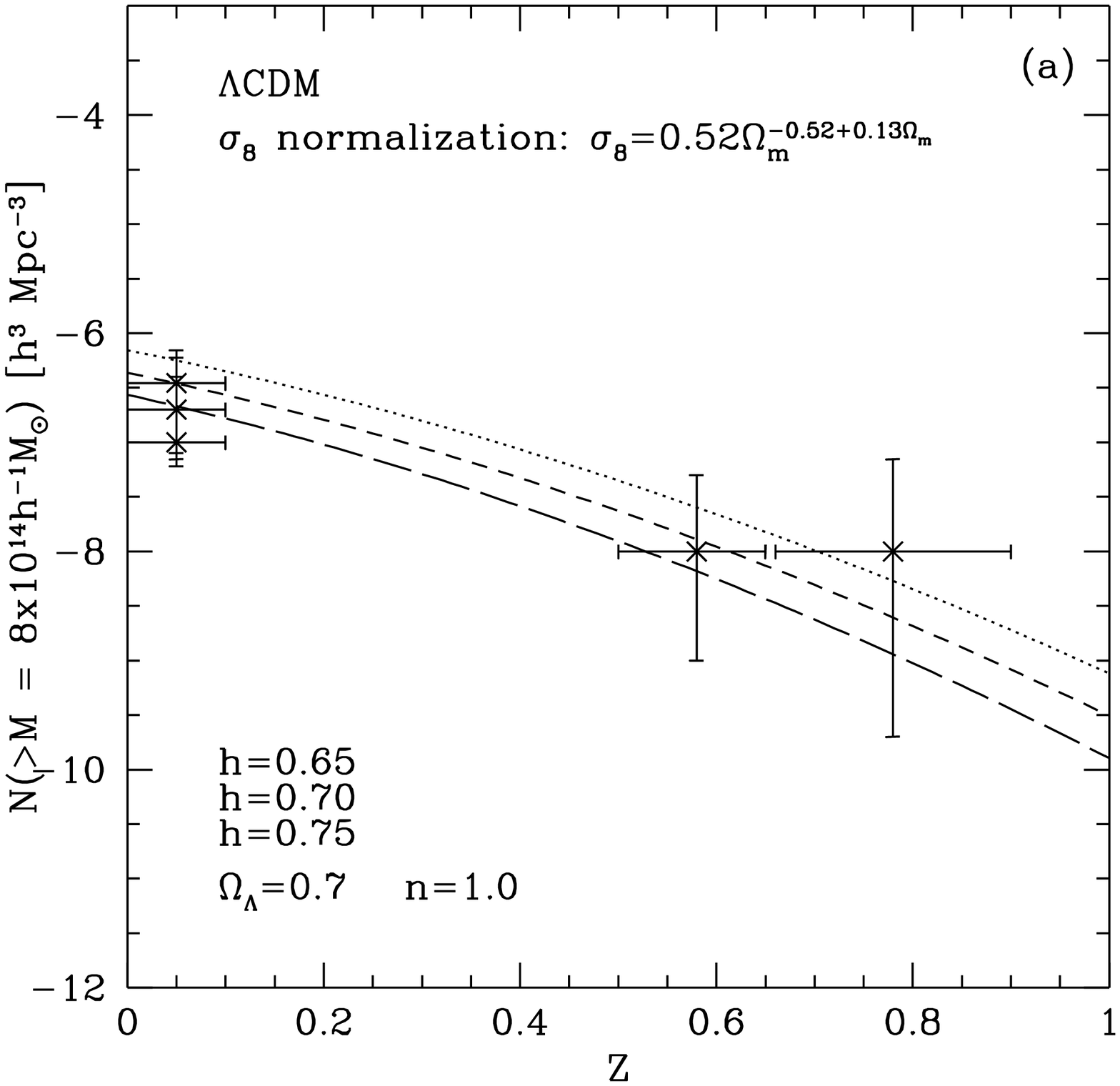}}
\resizebox{\hsize}{!}{\includegraphics{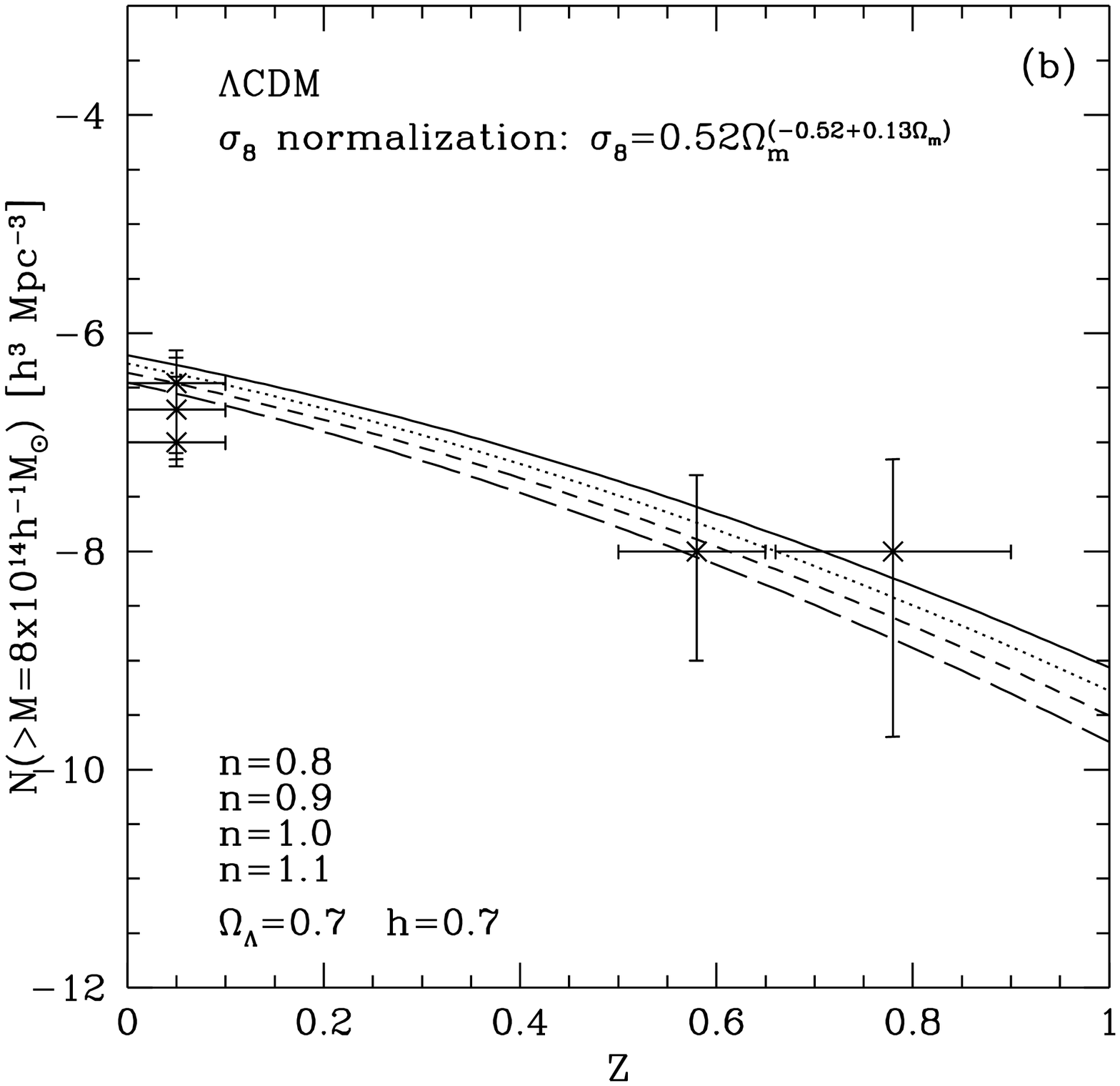}}             

\caption{
The cluster evolution $N(>M = 8 \cdot 10^{14} M_\odot, z)$
in $\Lambda$CDM models normalized by $\sigma_8=0.52 \Omega_m^{(-0.52+0.13\Omega_m)}$ with 
$\Omega_{\Lambda}=0.7$, $\Omega_b=0.015/h^2$, (a) $n=1$,
$h=0.65, 0.70, 0.75$ 
(dot, dash, long-dash lines, resp.), and 
(b) $h=0.7$,  $n=0.8, 0.9, 1.0, 1.1$ 
(solid, dot, dash, long-dash lines, resp.). 
The data points by Bahcall \& Fan (1998). } 
\end{figure}

\begin{figure}
\resizebox{\hsize}{!}{\includegraphics{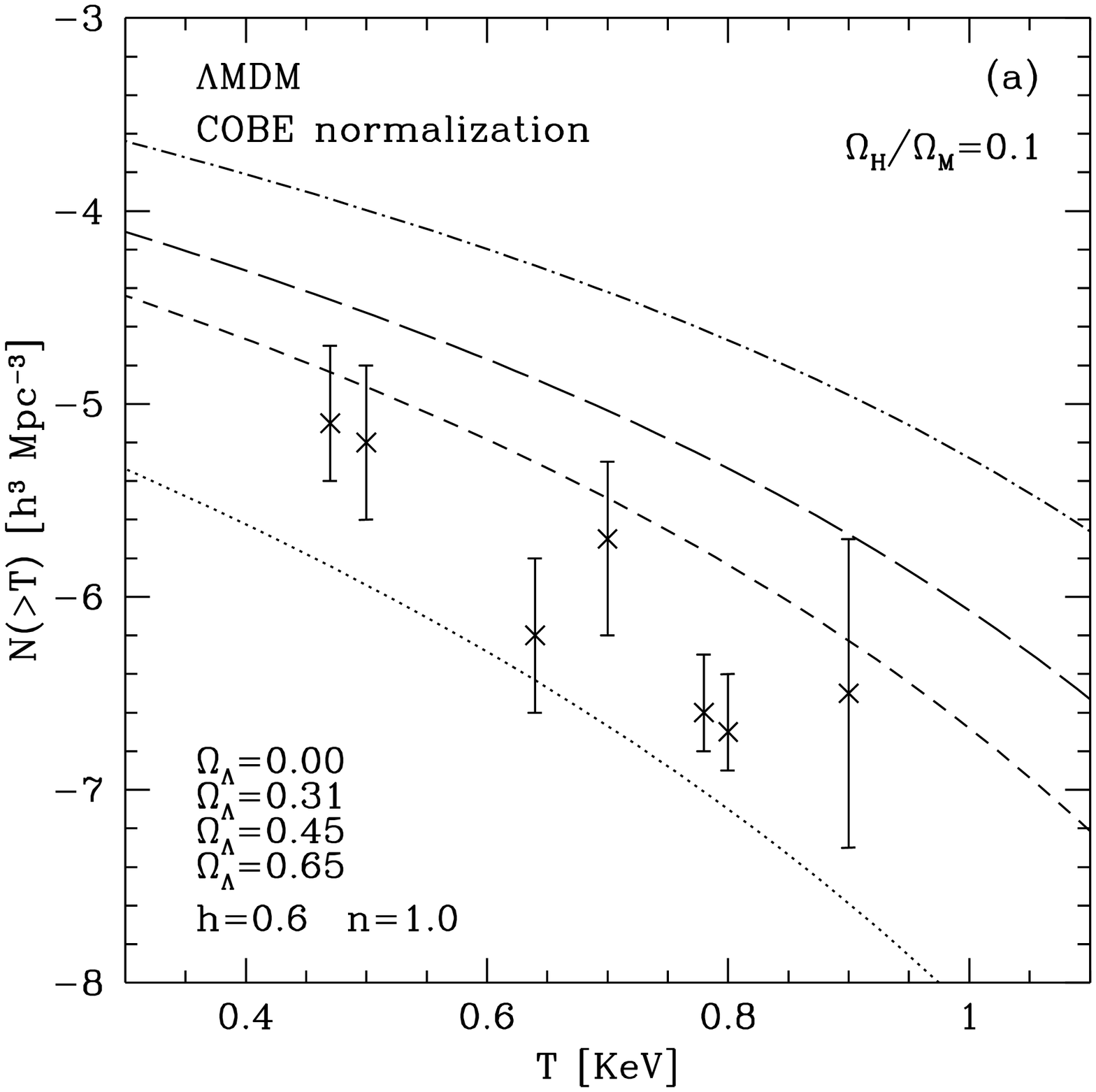}}
\resizebox{\hsize}{!}{\includegraphics{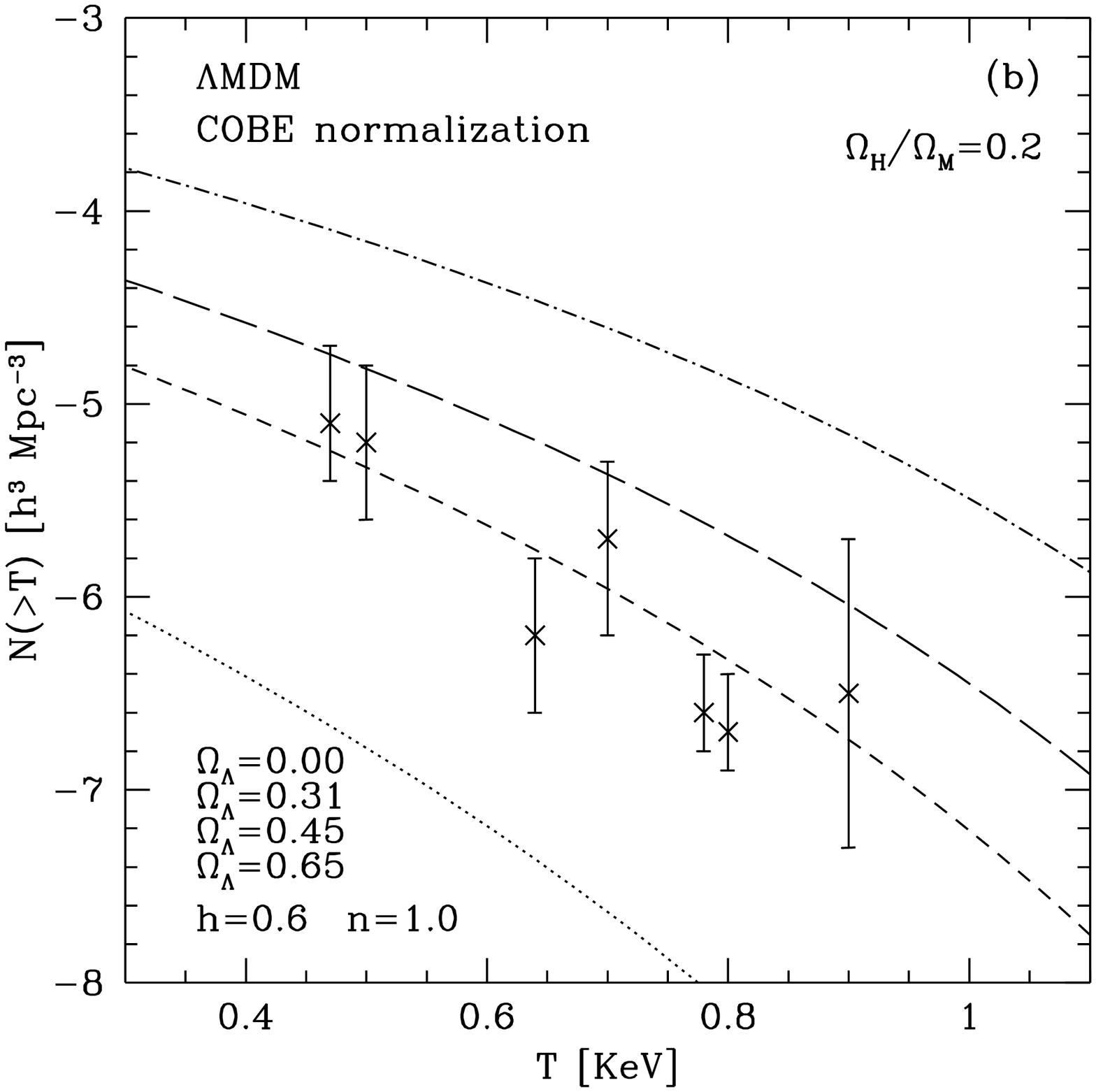}}      
\caption{The present day cluster abundance $N(>T)$ in $\Lambda$MDM
models normalized by COBE 4-year data with
 $\Omega_\Lambda=0, 0.31, 0.45, 0.65$ 
( dot-dash, long-dash, dash, dot lines, resp.),
$\Omega_b=0.015/h^2$, $h=0.6$, $n=1$, and
$f_\nu = 0.1$ (a), $0.2$ (b). 
The data points by Henry \& Arnaud (1991).}
\end{figure}

According to Henry (2000) the cluster evolution test has a stronger 
upper-limited parameter $\Omega_m$ in $\Lambda$CDM models than does 
the cluster abundance test at $z=0$. As an example 
we demonstrate the evolution of the number density of galaxy clusters of mass 
$M \geq 8 \cdot 10^{14} M_\odot$ within the comoving radius $1.5h^{-1}$Mpc 
in $\Lambda$CDM model with $\Omega_\Lambda=0.7$, as functions of $h$ 
(Fig.~4a) and $n$ (Fig.~4b).  The cluster evolution is not practically 
affected by changes of the spectral index. Regarding the parameter $h$, 
the models with $h > 0.6$ are preferable.

While the requested value of cosmological constant in $\Lambda$CDM models 
is quite high, a small amount of hot particles in $\Lambda$MDM models leads 
to the agreement with observations for a smaller $\Omega_\Lambda$. 

We performed the calculation of functions $N(>T)$ for different parameters
$\Omega_\Lambda$ and $f_\nu$ (Fig. 5). As we can see, to achieve an agreement 
with today cluster temperature function, we need a lower value of 
cosmological constant in the models with a higher fraction of hot matter.
For $f_\nu\in (0, 0.2)\;, h\in (0.6, 0.7)\;, n\in (0.9, 1.1)\;$, the value 
of cosmological constant remains in the range ($\;1\sigma$ CL):
\begin{equation}
0.35 <\Omega_\Lambda < 0.7\;.
\end{equation} 
Taking into account the $Ly_\alpha$ forest test, models with 
$f_\nu \simeq 0.1$ are preferable. In these case parameter $\Omega_\Lambda$ 
is limited by $0.45 \leq \Omega_\Lambda < 0.65$ (in agreement with some 
other results, cf. Valdarnini et al.(1998), Novosyadlyj et al.(1999b), 
Primack \& Gross (2000), Durrer \& Novosyadlyj (2001)).

\begin{figure}
\resizebox{\hsize}{!}{\includegraphics{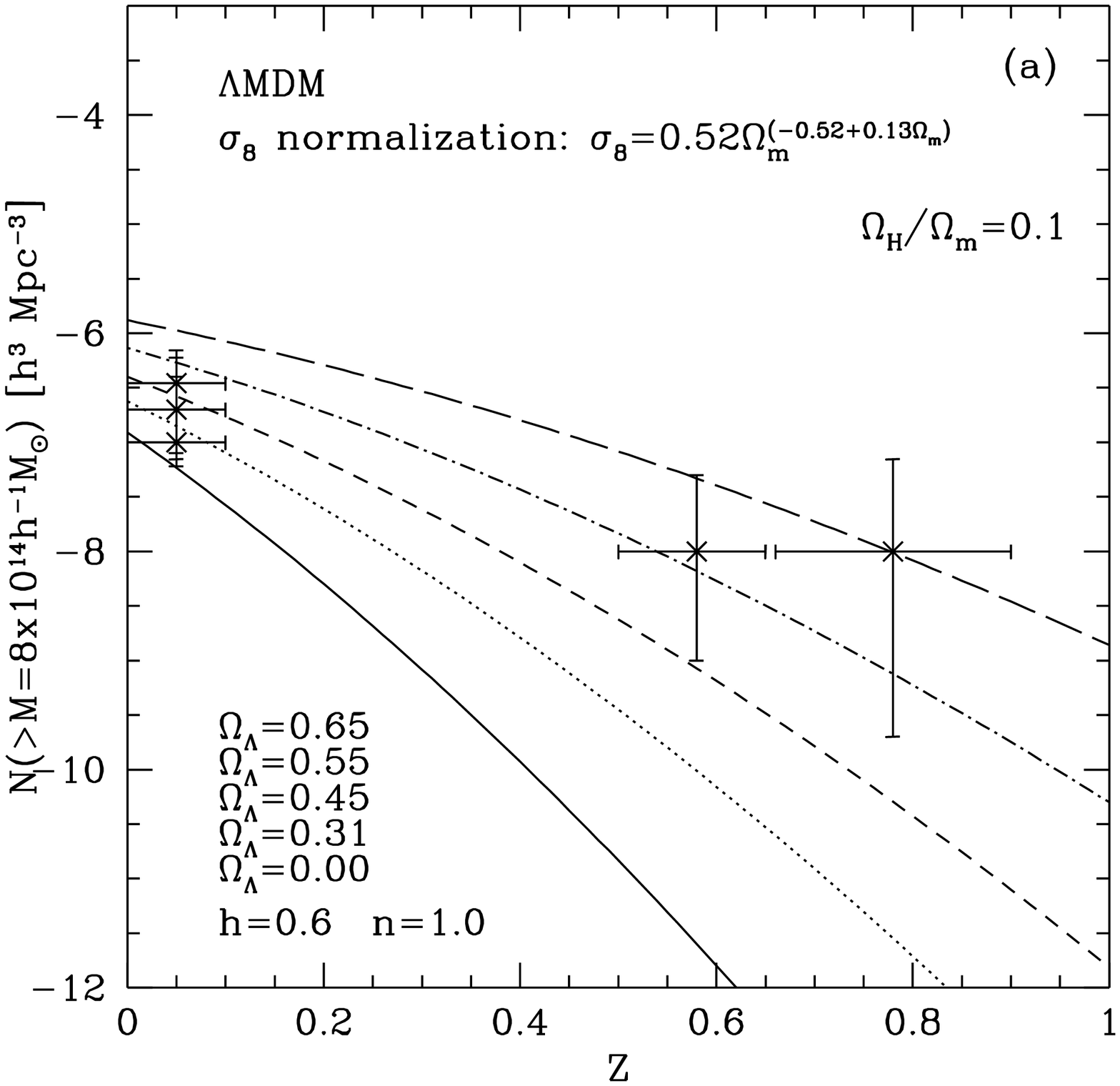}}
\resizebox{\hsize}{!}{\includegraphics{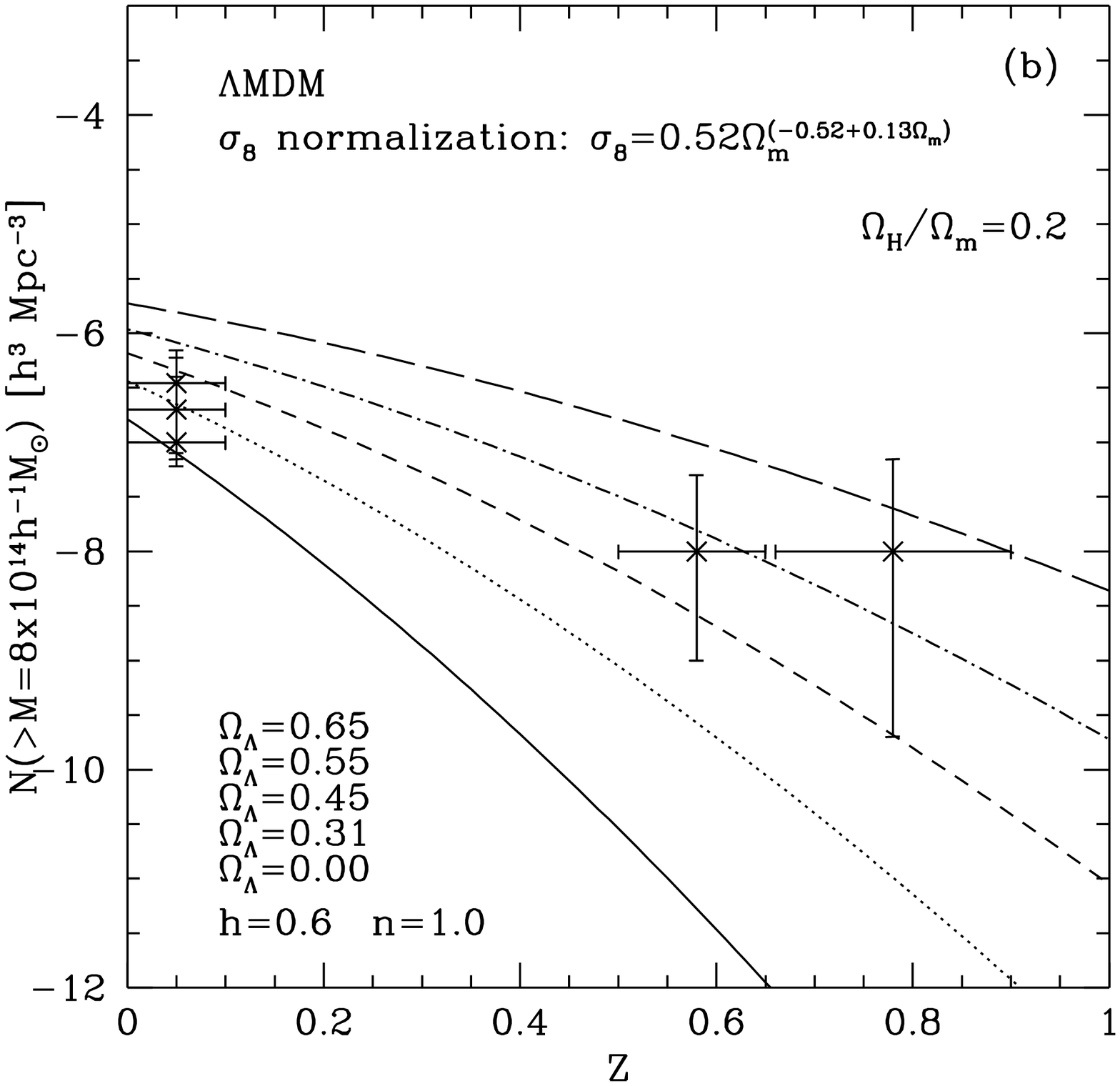}}     
\caption{The cluster evolution $N(>M = 8 \cdot 10^{14} M_\odot, z)$ 
in $\Lambda$MDM models normalized by 
$\sigma_8=0.52 \Omega_m^{(-0.52+0.13\Omega_m)}$ 
with  $\Omega_\Lambda =0, 0.31, 0.45, 0.55, 0.65$ 
(solid, dot, dash, dot-dash, long-dash lines, resp.), 
$\Omega_b=0.015/h^2$, $h=0.6$, $n=1$, and  
$f_\nu = 0.1$ (a), $0.2$ (b). 
The data points by Bahcall \& Fan (1998).}
\end{figure}

\begin{figure}
\resizebox{\hsize}{!}{\includegraphics{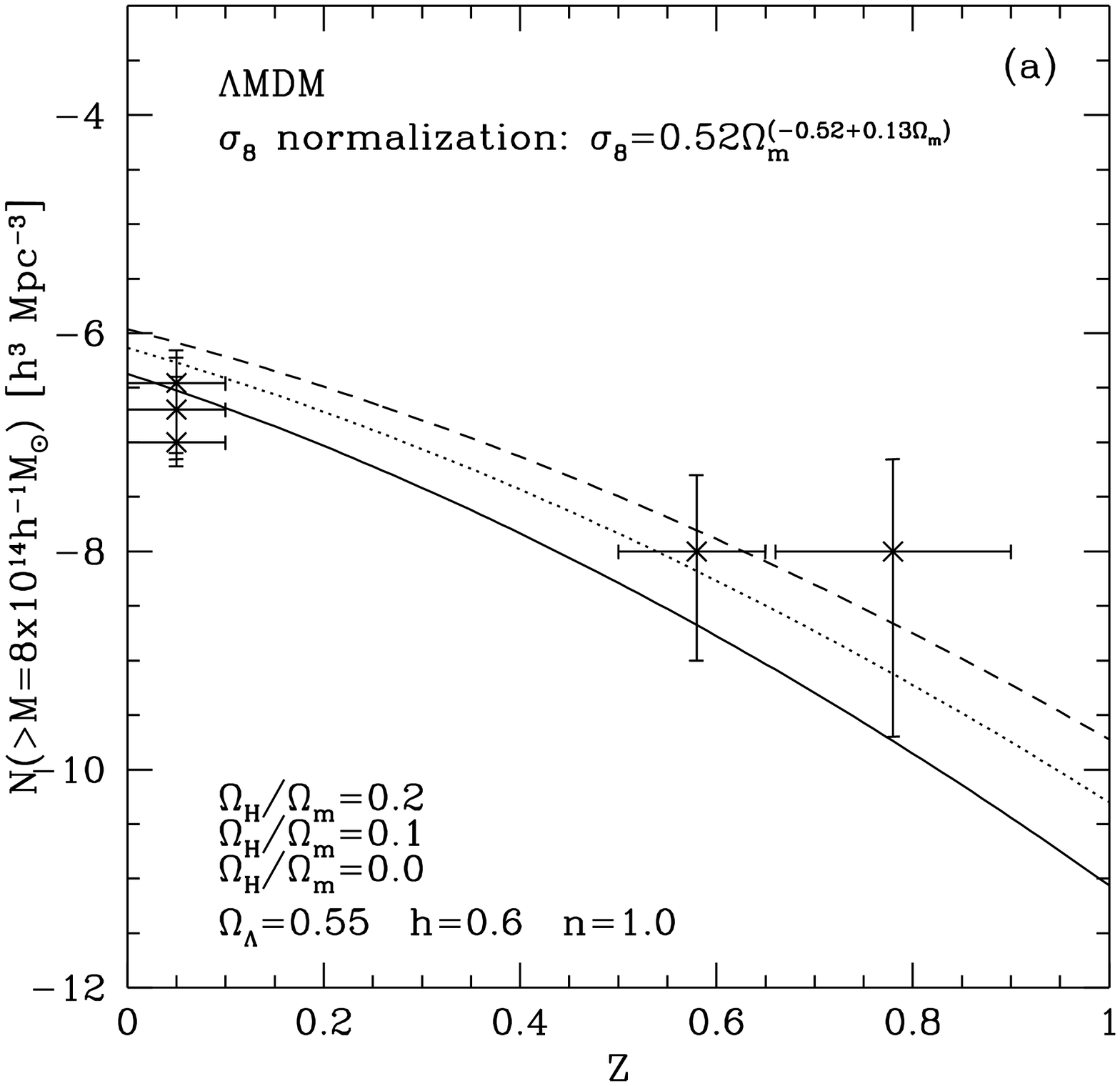}}
\resizebox{\hsize}{!}{\includegraphics{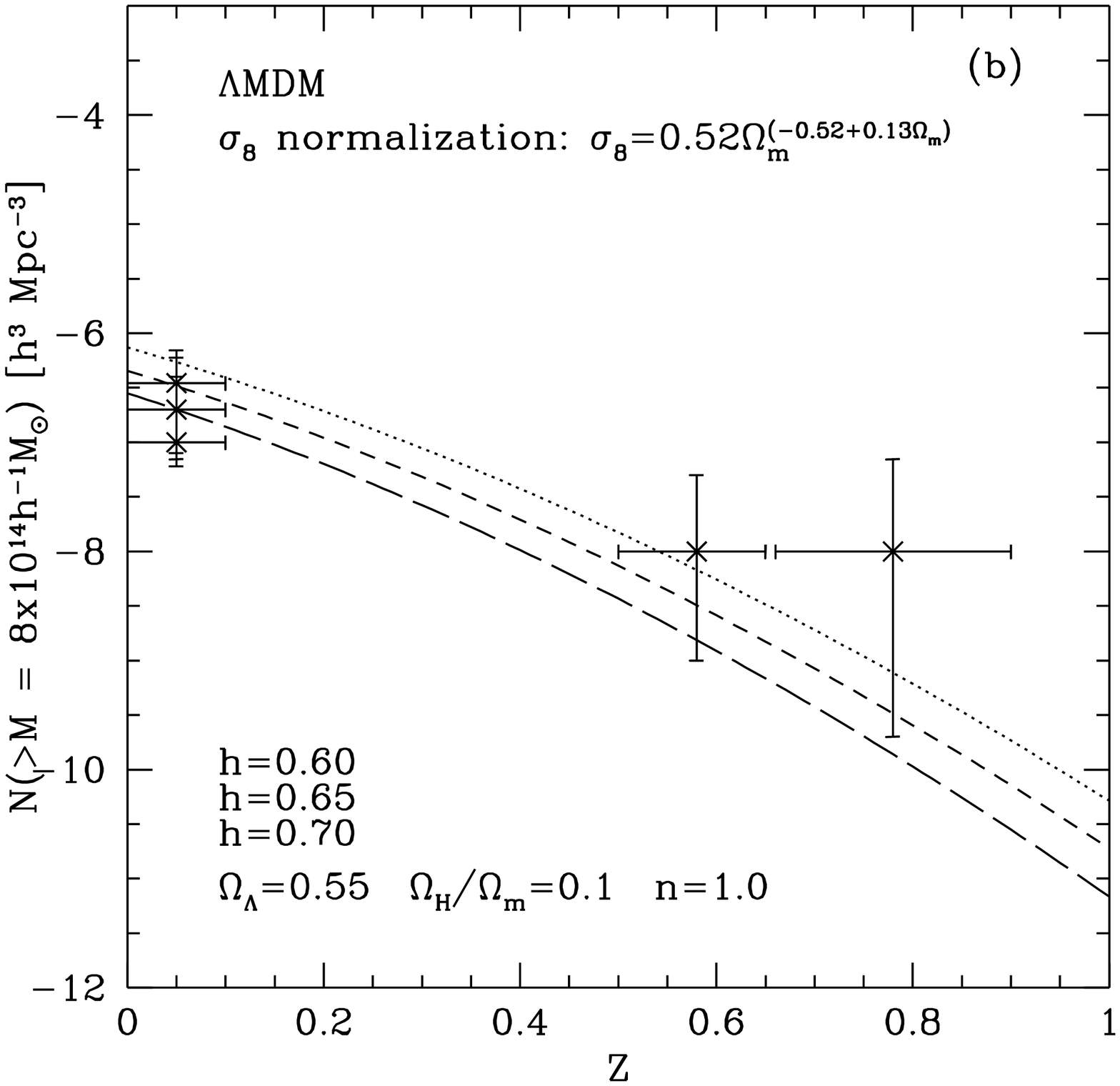}}
\caption{The cluster evolution $N(>M = 8 \cdot 10^{14} M_\odot, z)$
in $\Lambda$MDM models normalized by 
$\sigma_8=0.52 \Omega_m^{-0.52+0.13 \Omega_m}$ with 
$\Omega_\Lambda=0.55$, $\Omega_b=0.015/h^2$, $n=1$, 
(a) $h=0.6$,  $f_\nu = 0, 0.1, 0.2$ 
(solid, dot, dash lines, resp), and 
(b) $f_\nu = 0.1$, $h=0.6, 0.65, 0.7$ 
(dot, dash, long-dash lines, resp.).  
 The data points by Bahcall \& Fan (1998).}
\end{figure}

To obtain the redshift evolution functions of galaxy cluster abundance, 
the normalization of the power spectrum is done according
to the present-day cluster concentration, $\sigma_8(\mbox{cl})$, 
which depends on $\Omega_\Lambda$. At $1\sigma$ level the best 
fit for $\sigma_8(\mbox{cl})$ is not sensitive to the hot particle 
abundance, $h\in (0.6, 0.7)$, and $n\in (0.9, 1.1)$. 
Figs.~6,7 present the galaxy cluster evolution in $\Lambda$MDM models 
as function of parameters $\Omega_\Lambda$, $f_\nu$, and $h$.    
Changing $h$ and $n$ (within their ranges) does not practically influence 
the evolution tracks. Increasing $f_\nu$ reduces the needed value of 
$\Omega_\Lambda$ (in agreement with the cluster number density test).

For $h\in (0.6, 0.7)$, $n\in (0.9, 1.1)$, the value of $\Omega_\Lambda$ 
depends mainly on the fraction of hot matter $f_\nu$. 
We find the following approximation between the parameters $\Omega_\Lambda$ 
and $f_\nu \in (0, 0.2)$ from the cluster redshift evolution alone 
($1\sigma$ CL):
\begin{equation}
\Omega_\Lambda +0.5f_\nu =0.65\pm 0.1\;. 
\end{equation}
In connection with this degeneracy two points should be emphasized.
\begin{itemize}
\item[*]
As we see, the redshift evolution test slightly exaggerates the value of 
$\Omega_\Lambda$ in comparioson with the present day cluster abundance 
(cf. eq.(10)). To be in the agreement with both tests the value of
the cosmological constant in $\Lambda$MDM models should be found as 
$\Omega_\Lambda\in (0.45, 0.7)$.\\
\item[*]
Eq.(11) can also be understood as a proportionality between the values
of $\Omega_m$ and $f_\nu$: more matter leads to a greater fraction of hot 
matter and {\it{vice verce}}. Such a degeneracy is physically clear and could 
be straightforwardly observed in LSS tests. E.g., Novosyadlyj' et al. (1999b) 
indicate the correlation between  $\Omega_m$ and $\Omega_\nu$ from the 
cluster power spectrum and $Ly_\alpha$ data. We guess that a joint LSS 
analysis will confirm eq.(11) with a better precision.
\end{itemize}
In conclusion, we stress that the present observational data constrain 
only the parameter $f_\nu \in (0, 0.2)$, the number of species 
of massive neutrino ($N_\nu= 1,2,3\;$) remains undetermined.
  
\section{Conclusions}

Our main conclusions are as follows:
\begin{itemize}
\item
The introduction of massive neutrino in spatially flat 
dark matter models affects drastically the estimation of the 
cosmological parameters from galaxy cluster observational data.\\
\item
MDM models with cosmic gravitational waves and a negligible
$\Lambda$-term do not pass the observational constraints: 
they require a significal contribution of both 
(i) CGW (T/S$\;{}^>_\sim 1$) and 
(ii) hot matter ($f_\nu{}^>_\sim 0.2$). 
The first condition reduces the first acoustic peak in 
$\Delta T/T$ spectrum by a level inconsistent with 
the BOOMERANG measurements, the second  condition 
contradicts the $Ly_\alpha$-system formation test.\\
\item
The introduction of even a small cosmological constant in 
MDM-CGW models essentially improves the situation with the
model parameters, indicating that $\Lambda$-term is a more 
powerful instrument (than T/S) to reconcile theory with observations.\\
\item
The cluster abundance and/or evolution tests alone clearly 
hint at the existence of a non-zero cosmological constant in 
a set of spatially flat $\Lambda$MDM models with 
a negligible amount of CGW. A rise of T/S cannot cancel the $\Lambda$-term. 
However, the opposite is not true: the estimation of T/S in 
$\Lambda$MDM cosmologies requires better data and a more subtle analysis.\\
\item
The presence of a small fraction of massive neutrino, 
$f_\nu\in (0, 0.2)\;$, (i) reduces the required value 
of cosmological constant (in comparison with $\Lambda$CDM), 
(ii) is preferable from the point of view of the $ly_\alpha$ 
forest test, and (iii) does not discriminate the
estimate of other cosmological parameters, $h\in (0.6, 0.7)$, 
$n\in (0.9, 1,1)$, $N_\nu = 1, 2, 3\;$.\\ 
\item
For $f_\nu\in (0, 0.2),\; h\in (0.6, 0.7),\; n\in (0.9, 1.1)$, 
the value of cosmological constant remains in $1\sigma$ ranges:\\
$*\;\; \Omega_\Lambda\in (0.35, 0.7)\;$ (from nearby cluster abundance),
\\
$*\;\; \Omega_\Lambda\in (0.45, 0.75)\;$ (from cluster redshift evolution),\\ 
$*\;\; \Omega_\Lambda\in (0.45, 0.7)\;$ (from the both tests).\\
\item
We find the following approximation between parameters $\Omega_\Lambda$ 
and $f_\nu\in (0, 0.2)$ from the cluster redshift evolution alone:
\[
\Omega_\Lambda +0.5f_\nu =0.65\pm 0.1\;\;,\;\; (1\sigma \rm{CL}).
\]
\item
If massive neutrinos constitute only $\sim 10\%$ of the total 
DM density ($f_\nu = 0.1$) the models with 
$\Omega_\Lambda\in (0.5, 0.65)$ satisfy the observational data best.  
\end{itemize}

It may well be that the DM nature is more complex than just $\Lambda$MDM 
with a negligible amount of CGW. Moreover, even in the framework of 
this simple model we cannot delimit today the fraction 
and the number of species of massive neutrino, the accuracy of the 
available data is still low to do it. Nevertheless, 
we can assert that the abundance of cosmological massive neutrinos is small, 
$f_\nu < 0.2$, corresponding less than few eV to the neutrino mass. 
Also, today we can conclude that
\begin{itemize}
\item[*] 
the value of the estimated cosmological constant is very sensitive 
to even a small fraction of massive neutrinos, and\\
\item[*]
the continuing progress in LSS cosmological observations 
assists in solving the DM problem.
\end{itemize}

{\it Acknowledgments}
The work of N.A.A. and V.N.L. was partly supported by RFBR (01-02-16274, 
00-15-96639), INTAS (97-1192), and SST Program in Astronomy ("kosmomikrofizika"). 
The work of T.K. was supported by the COBASE program of USNRC.

\end{document}